\documentclass[11pt,letterpaper]{amsart}

\usepackage{amssymb}

\newtheorem{Thm}{Theorem}

\newcommand{\Z}{\ensuremath{\mathbb{Z}}}

\newcommand{\be}{\begin{eqnarray}}
\newcommand{\ee}{\end{eqnarray}}
\newcommand{\bes}{\begin{eqnarray*}}
\newcommand{\ees}{\end{eqnarray*}}

\newcommand{\ostar}[0]{\ensuremath{\circledast}}
\newcommand{\covprod}[0]{\ensuremath{*_\mathrm{c}}}
\newcommand{\covprodi}[0]{\ensuremath{*_\mathrm{ic}}}
\newcommand{\packprod}[0]{\ensuremath{*_\mathrm{p}}}

\begin{document}

%%%%%%%%%%%%%%%%%%%%%%%%%%%%%%%%%%%%%%%%%%%%%%%%%% Front matter information %%%

\title%
[]%
{Fourier Meets M\"obius: Fast Subset Convolution}
\author[]{\vspace*{-0.1in}Andreas Bj\"{o}rklund}
\address{%
Lund University, 
Department of Computer Science,  
P.O.Box 118, SE-22100 Lund, Sweden}
\email{andreas.bjorklund@anoto.com}
\author[]{Thore Husfeldt}
\address{%
Lund University, 
Department of Computer Science,  
P.O.Box 118, SE-22100 Lund, Sweden}
\email{thore.husfeldt@cs.lu.se}
\author[]{Petteri Kaski}
\address{%
Helsinki Institute for Information Technology HIIT,
Department of Computer Science, University of Helsinki,
P.O.Box 68, FI-00014 University of Helsinki, Finland}
\email{petteri.kaski@cs.helsinki.fi}
\author[]{Mikko Koivisto}
\address{%
Helsinki Institute for Information Technology HIIT,
Department of Computer Science, University of Helsinki,
P.O.Box 68, FI-00014 University of Helsinki, Finland}
\email{mikko.koivisto@cs.helsinki.fi}

\newcommand{\abst}[0]{%
We present a fast algorithm for the {\em subset convolution problem}:
given functions $f$ and $g$  defined on the lattice of subsets of an
$n$-element set $N$, compute their {\em subset convolution} $f*g$, 
defined for all $S\subseteq N$ by  
\bes
	(f * g)(S) = \sum_{T \subseteq S} f(T) g(S\setminus T)\,,
\ees
where addition and multiplication is carried out in an  arbitrary
ring. Via M\"{o}bius transform and inversion, our algorithm  
evaluates the subset convolution in $O(n^2 2^n)$ additions and
multiplications, substantially improving upon the straightforward
$O(3^n)$ algorithm. Specifically, if the input functions have an
integer range $\{-M,-M+1,\ldots,M\}$, their subset convolution over
the ordinary sum--product ring can be computed in $O^*(2^n \log M)$
time;  the notation $O^*$ suppresses polylogarithmic factors.
Furthermore, using a standard embedding technique we can compute the
subset convolution over the max--sum or min--sum {\em semiring} in
$O^*(2^n M)$ time.

To demonstrate the applicability of fast subset convolution, we
present the first $O^*(2^k n^2 + n m)$ algorithm for the minimum
Steiner tree problem  in graphs with $n$ vertices, $k$ terminals, and
$m$ edges  with bounded integer weights,  improving upon the $O^*(3^k
n + 2^k n^2 + n m)$ time bound of the classical Dreyfus--Wagner
algorithm. We also discuss extensions to recent $O^*(2^n)$-time
algorithms for covering and partitioning problems  (Bj\"{o}rklund and
Husfeldt, FOCS 2006; Koivisto, FOCS 2006); using fast subset
convolution we can, for example, find {\em all} $k$-colorable induced
subgraphs of a given $n$-vertex graph in $O^*(2^n)$ time.
}

%%%%%%%%%%%%%%%%%%%%%%%%%%%%%%%%%%%%%%%%%%%%%%%%%%%%%%%% Title and abstract %%%

\begin{abstract}
\abst
\end{abstract}
\vspace*{-0.5in}
\maketitle
\vspace*{-0.2in}

%%%%%%%%%%%%%%%%%%%%%%%%%%%%%%%%%%%%%%%%%%%%%%%%%%%%%%%%%%%%% Document body %%%

\section{Introduction}

\subsection{Background and Main Result}
Many hard computational problems admit a recursive solution via
a convolution-like recursion step over the subsets of an $n$-element 
ground set $N$. More precisely, for every $S \subseteq N$, 
one computes the ``solution'' $h(S)$ defined by 
\begin{equation}
\label{eq:sc}
  h(S) = \sum_{T \subseteq S} f(T) g(S\setminus T)\,, 
\end{equation}
where $f(T)$ and $g(S\setminus T)$ are previously computed
solutions  for the subproblems specified by $T$ and
$S\setminus T$, and the arithmetic is carried out in an
appropriate semiring; the most common examples in
applications being perhaps the integer sum--product  ring
and the integer max--sum semiring. Given $f$ and $g$, a
direct evaluation of $h$ for all $S\subseteq N$  requires
$\Omega(3^n)$ semiring operations. To our knowledge, this is
also the fastest known evaluation approach until the present
work.

In a first attempt to improve upon the direct evaluation, 
the convolution analogy suggests the natural approach to 
evaluate (\ref{eq:sc}) as a product of some type of Fourier 
transforms of $f$ and $g$ via a fast Fourier transform 
(FFT) and its inverse---in general, this approach has proven to be 
spectacularly successful in domains ranging from signal processing 
to number theory; see \cite{Maslen_and_Rockmore_1997} for a survey of 
generalized FFTs. 
For example, considering a slightly different convolution operation 
of the form
\begin{equation}
\label{eq:sym-sc}
  h'(S) = 
  \sum_{T \subseteq N} 
    f\bigl(T\bigr) g\bigl(S\Delta T\bigr)\,,
	\qquad \textrm{where } 
  S\Delta T=(S\setminus T)\cup(T\setminus S)\,,
\end{equation}
the Fourier approach immediately yields an evaluation approach requiring 
$O(n2^n)$ ring operations via the fast Fourier transform on
$\Z_2^n$, the elementary Abelian group of order $2^n$.
However, the convolution (\ref{eq:sc}) is ``truncated'' 
from $T\subseteq N$ to $T\subseteq S$, which in effect renders 
the operation somewhat incompatible with group-theoretic Fourier 
transforms and associated ``natural'' convolution operations 
performed over the entire group. 
A simple zero-padding trick allows one to evaluate (\ref{eq:sc}) via
the FFT on $\Z_3^n$ (equivalently, the classical $n$-dimensional 
FFT with padding on each dimension), but this unfortunately does not 
improve upon the direct evaluation strategy.

A second attempt at analogy will prove to be more successful.
Indeed, the truncation to $T\subseteq S$ and the sum over all 
$T\subseteq S$ in (\ref{eq:sc}) suggests a connection with the classical 
M\"obius transform \cite{Aigner_1979,Rota_1964,Stanley_1997} on 
the lattice of subsets of $N$, which is further motivated by the fact 
that a fast algorithm is known for evaluating the M\"obius transform 
and its inverse on the subset lattice. 

It turns out  that---in analogy with the Fourier approach---the 
evaluation of (\ref{eq:sc}) in $O(n^22^n)$ ring operations can be 
achieved via a product (``convolution over rank'') of ``ranked'' 
extensions of the classical M\"obius transforms of $f$ and $g$ on the
subset lattice, followed by a ``ranked'' M\"obius inversion.  This is
the main result of this paper.

\begin{Thm}
\label{thm:main}
The subset convolution over an arbitrary ring can be 
evaluated in $O(n^22^n)$ ring operations.
\end{Thm}

Furthermore, considerable extensions and variations of the basic fast 
convolution operation are possible, which we expect to find applications
beyond the ones we proceed to outline in what follows.

\subsection{Application to Specific Computational Problems}

Besides the algebraic complexity we also study the
implementation of the fast subset convolution algorithm on
the ordinary sum--product ring of integers. The model of
computation assumed in our analyses is the random access
machine with the restriction that arithmetic operations
(including comparison) are considered unit-time only for
integers of constant size. 
To avoid cumbersome expressions in runtime bounds we may use the
notation $O^*$ to hide polylogarithmic factors, that is, we
may denote $O^*(\tau)$ when we have $O(\tau \log^d \tau)$ for some
constant $d$ in the familiar Landau notation. 

Our main result easily implies the following. 

\begin{Thm}\sloppy
\label{thm:intsumprod}
The subset convolution 
over the integer sum--product ring 
can be computed in $O^*(2^n \log M)$
time, provided that the range of the input functions is\/
$\{-M,-M+1,\ldots,M\}$. 
\end{Thm}

Combinatorial optimization problems usually concern the
max--sum or min--sum {\em semiring}. While our fast subset
convolution  algorithm does not directly apply to semirings
where additive inverses need not exist, we can, fortunately,
embed the integer max--sum (min--sum) semiring into the
integer sum--product ring. 

\begin{Thm}\sloppy
\label{thm:intminsum}
The subset convolution
over the integer max--sum $($\hspace*{-1pt}min--sum$)$ semiring 
can be computed in $O^*(2^n M)$ time, 
provided that the range of the input functions is\/
$\{-M,-M+1,\ldots,M\}$.
\end{Thm}

As an illustrative application of fast subset convolution, we  
accelerate the classical Dreyfus--Wagner algorithm
\cite{Dreyfus_and_Wagner_1972} for the minimum Steiner tree problem:
given an undirected graph $G = (V, E)$,  a weight $w(e) > 0$ for each
edge $e \in E$, and a set of vertices $K\subseteq V$, find a
minimum-weight subgraph
$H$ of $G$ that connects the vertices in $K$.  The
Dreyfus--Wagner algorithm runs in  $O^*(3^k n + 2^k n^2 + n m)$ time,
where $n = |V|$, $m = |E|$, and $k=|K|$. We give the first $O^*(2^k
n^2 + n m)$-time algorithm, provided that the edge weights are small
integers.

The Dreyfus--Wagner algorithm and its variants play a key role  in
solving various related problems.  For example, the Dreyfus--Wagner
algorithm has recently been used  as a subroutine in fixed parameter
tractable algorithms for  certain vertex cover problems
\cite{Guo_etal_2005}, as well as for  near-perfect phylogenetic tree
reconstruction \cite{Blelloch_etal_2006}. Regarding rectilinear
Steiner trees (RSTs),  Ganley \cite{Ganley_1999} writes: ``The
algorithm of Dreyfus and Wagner is probably the most popular used to
date for computing optimal RSTs in practice.''  Furthermore, other
``hierarchical partitioning'' algorithms similar to that of Dreyfus
and Wagner seem to appear in the literature  with no explicit
connection to the minimum Steiner tree problem;  we will consider in
some detail a recent algorithm by Scott, Ideker, Karp, and Sharan
\cite{Scott_etal_2005} for detecting signaling pathways in protein
interaction networks. Our improvement via fast subset convolution 
concerns all these variants and applications, subject to the
constraint that edge weights can be represented by small integers.

We also note that many classical graph partitioning problems
\cite{Garey_and_Johnson_1979} can be solved by counting all  valid
partitions via recursive application of subset convolution  (over the
integer sum--product ring).  Thus, the present technique can be seen
as a  generalization of the authors' previous work
\cite{Bjorklund_and_Husfeldt_2006, Koivisto_2006} based on
inclusion--exclusion  and applies to a wider family of partitioning
problems.  For example, we can now solve extended partitioning
problems, such as   finding {\em all} $k$-colorable induced subgraphs
of a given $n$-vertex graph, in $O^*(2^n)$ total time.

\subsection{Related Research and Discussion}

M\"{o}bius transform and inversion play a central role in
combinatorial theory, particularly in the theory of
partially ordered sets, subset lattices being special cases 
\cite{Aigner_1979,Rota_1964,Stanley_1997}.
The fast M\"{o}bius transform and inversion algorithms on
the subset lattice can be considered folklore; Kennes
\cite{Kennes_1991} gives a formal treatment, but the
algorithm is, in essence, that of Yates \cite{Yates_1937}
for multiplying a vector of size $2^n$ by a Kronecker
product of $n$ matrices of size $2 \times 2$ in $O(n 2^n)$
operations. 
As far as we know, the connection of (ranked) 
M\"{o}bius inversion and subset convolution has not
been studied until the present work. 

For partitioning problems, we (the first two authors and the
last author, independently) recently found two mutually
different inclusion--exclusion algorithms
\cite{Bjorklund_and_Husfeldt_2006, Koivisto_2006}, which
anticipated the results of the present paper. Yet, these
earlier results, even when combined, do not immediately
yield the fast subset convolution algorithm. What remained
to be discovered was, in essence, the role of (fast) ranked 
M\"{o}bius inversion. 

For the minimum Steiner tree problem,   M\"{o}lle, Richter, and
Rossmanith \cite{Molle_etal_2006} have recently found an  algorithm
that, for any fixed $\epsilon > 0$, runs in  $O^*((2+\epsilon)^k
p(n))$ time, where $p(n)$ is a polynomial  function of $n$.
Unfortunately, the degree of $p(n)$ grows rapidly  when $\epsilon$
approaches zero, which renders the algorithm impractical for small
$\epsilon$; in a subsequent work \cite{Fuchs_etal_2006}, the degree of
$p(n)$ is improved to $12 \sqrt{\epsilon^{-1}\ln \epsilon^{-1}}$, 
resulting in bounds like $O^*(2.5^k n^{14.2})$ and $O^*(2.1^k n^{57.6})$. 
Our accelerated Dreyfus--Wagner algorithm is not only
theoretically faster, but may also have practical value when $k$ is
large enough (say, $k > 25$).  

The idea of embedding the integer max--sum or min-sum semiring into the
sum--product ring is not new. Some well-known examples are
Yuval's \cite{Yuval_1976} and others'
\cite{Galil_and_Margalit_1997, Seidel_1995, Shoshan_and_Zwick_1999} 
approaches to compute shortest paths via (fast) matrix
multiplication. 
In our case the embedding technique provides a more
substantial gain. Indeed, compared to fast subset convolution, 
fast matrix multiplication algorithms involve large constant factors
and their practical value is not clear; the exponential
speedup offered by fast matrix multiplication is currently
by the ratio $3/2.376$ \cite{Coppersmith_and_Winograd_1990} 
and cannot exceed $3/2$,
whereas the exponential speedup offered by fast subset
convolution is by the ratio $\log 3 / \log 2 > 1.58 > 3/2$. 

\subsection{Organization}

The next section is devoted to proving our main theorem
(Theorem~\ref{thm:main}); in addition, we introduce some variants of
the subset convolution problem together with corresponding fast
algorithms.  In Section~3 we give short proofs of 
Theorems~\ref{thm:intsumprod} and \ref{thm:intminsum} concerning  the
implementation in the sum--product ring and the max--sum and  min--sum
semirings, which are essential for the applications. We consider the
minimum Steiner tree problem in detail in Section~4; also other
applications  to extended partitioning and hypergraph problems are
illustrated, but in somewhat less detail.

\section{Fast Subset Convolution over a Ring}

Throughout this section we assume that $R$ is 
an arbitrary (possibly noncommutative) ring
and that $N$ is a set of $n$ elements, $n\geq 0$.

\subsection{Subset Convolution}
Let $f$ be a function that associates with every subset $S\subseteq N$
an element $f(S)$ of the ring $R$. 
For two such functions, $f$ and $g$, define the 
\emph{convolution} $f*g$ for all $S\subseteq N$ by
\begin{equation}
\label{eq:convolution}
	(f*g)(S)
	=\sum_{T\subseteq S} f(T)g(S\setminus T)\,,
\end{equation}
or, equivalently, in a more symmetric form 
\begin{equation}
\label{eq:convolution2}
	(f*g)(S)
	=\sum_{\tiny\begin{array}{c}U,V\subseteq S\\U\cup V=S\\U\cap V=\emptyset\end{array}} f(U)g(V)\,.
\end{equation}
It follows that the convolution operation is associative
(and commutative if $R$ is commutative).

\subsection{M\"obius Transform and Inversion on the Subset Lattice}
We recall the classical M\"obius transform and inversion formulas
on the subset lattice together with their fast evaluation algorithms.
Let $f$ be a function that associates with every subset $S\subseteq N$
an element $f(S)$ of the ring $R$. 
The \emph{M\"obius transform} of $f$ is the function
$\hat f$ that associates with every $X\subseteq N$ the
ring element
\begin{equation}
\label{eq:moebius-trans}
	\hat f(X)=\sum_{S\subseteq X}f(S)\,.
\end{equation}
Given the M\"obius transform $\hat f$, the original function 
$f$ may be recovered via the \emph{M\"obius inversion} formula
\begin{equation}
\label{eq:moebius-inv}
	f(S)=\sum_{X\subseteq S}(-1)^{|S\setminus X|}\hat f(X)\,.
\end{equation}

The M\"obius transform (\ref{eq:moebius-trans})
can be computed in $O(n2^n)$ ring operations, which
constitutes the \emph{fast M\"obius transform}.
By relabeling if necessary, we may assume that
$N=\{1,2,\ldots,n\}$. To compute $\hat f$ given $f$, let initially
\[
	\hat f_0(X)=f(X)
\]
for all $X\subseteq N$, and then iterate 
for all $j=1,2,\ldots,n$ and $X\subseteq N$ as follows:
\begin{equation}
\label{eq:moebius-trans-fast}
	\hat f_j(X)=
	\begin{cases}
	\hat f_{j-1}(X) & \text{if $j\notin X$}\,,\\
	\hat f_{j-1}(X\setminus\{j\})+\hat f_{j-1}(X)  & \text{if
	$j\in X$}\,.
\end{cases}
\end{equation}
It is straightforward to verify by induction on $j$ that this recurrence
gives $\hat f_n(X)=\hat f(X)$ for all $X\subseteq N$ in $O(n 2^n)$ 
ring operations. The inversion operation (\ref{eq:moebius-inv}) 
can be implemented in a similar fashion.
To compute $f$ given $\hat f$, let initially
\[
f_0(S)=\hat f(S)\,
\]
for all $S\subseteq N$, and then iterate 
for all $j=1,2,\ldots,n$ and $S\subseteq N$ as follows: 
\begin{equation}
\label{eq:moebius-inv-fast}
f_j(S)=
\begin{cases}
 f_{j-1}(S) & \text{if $j\notin S$}\,,\\
-f_{j-1}(S\setminus\{j\})+f_{j-1}(S) & \text{if $j\in S$}\,.
\end{cases}
\end{equation}
Then we have $f_n(S) = f(S)$ for all $S \subseteq N$.

\subsection{Ranked M\"obius Transform and Inversion}
Let $f$ be a function that associates with every subset $S\subseteq N$
an element $f(S)$ of the ring $R$. 
The \emph{ranked M\"obius transform} of $f$ is the 
function $\hat f$ that associates with every $k=0,1,\ldots,n$ and
$X\subseteq N$ the ring element
\begin{equation}
\label{eq:rank-moebius-trans}
	\hat f(k,X)=\sum_{\tiny\begin{array}{c}S\subseteq X\\|S|=k\end{array}}f(S)\,.
\end{equation}
In particular, the classical M\"obius transform of $f$ is
obtained in terms of the ranked transform by taking the 
sum over $k$, that is, $\hat f(X)=\sum_{k=0}^{|X|}\hat f(k,X)$.
For the ranked transform, inversion is achieved simply by
\begin{equation}
\label{eq:rank-moebius-inv2}
	f(S)=\hat f(|S|,S)\,,
\end{equation}
or, in a somewhat more redundant form,
\begin{equation}
\label{eq:rank-moebius-inv}
	f(S)=\sum_{X\subseteq S}(-1)^{|S\setminus X|}\hat f(|S|,X)\,.
\end{equation}
This latter expression, rather than the former one, 
provides the key to fast evaluation of
the subset convolution (\ref{eq:convolution}). Namely, we will
``invert'' a function that, in general, cannot be 
represented via ranked M\"{o}bius transform but via a
convolution (over rank) of two such transforms.

To set the stage, it is immediate that the ranked transform 
(\ref{eq:rank-moebius-trans}) can be computed in $O(n^22^n)$ ring 
operations by carrying out the standard fast transform 
(\ref{eq:moebius-trans-fast}) independently for each 
$k=0,1,\ldots,n$. Similarly, the ranked
inversion (\ref{eq:rank-moebius-inv}) can be computed
in $O(n^22^n)$ ring operations by carrying out the standard fast 
inversion (\ref{eq:moebius-inv-fast}) independently for 
each $k=0,1,\ldots,n$.

\subsection{Fast Subset Convolution}
For two ranked M\"obius transforms, 
$\hat f$ and $\hat g$, define the convolution
$\hat f\ostar\hat g$ for all $k=0,1,\ldots,n$ and $X\subseteq N$ by
\begin{equation}
\label{eq:rank-convolution}
	(\hat f\ostar\hat g)(k,X)=\sum_{j=0}^k \hat f(j,X)\hat g(k-j,X)\,.
\end{equation}
Note that this convolution operation is over the rank parameter
rather than over the subset parameter.

It now holds that the inversion operation (\ref{eq:rank-moebius-inv})
applied to $\hat f\ostar\hat g$ gives $f*g$. 
Indeed, first observe by (\ref{eq:rank-moebius-trans}) 
and (\ref{eq:rank-convolution}) that for any $S\subseteq N$ we have
\begin{equation}
\label{eq:rank-convolution-inv}
\begin{split}
\sum_{X\subseteq S}
(-1)^{|S\setminus X|}(\hat f\ostar\hat g)(|S|,X)&=
\sum_{X\subseteq S}\ 
(-1)^{|S\setminus X|}\ \sum_{j=0}^{|S|}\  \hat f(j,X)\hat g(k-j,X)\\
&=\sum_{X\subseteq S}\ 
(-1)^{|S\setminus X|}\ \sum_{j=0}^{|S|}\!\!
\sum_{\tiny\begin{array}{c}U,V\subseteq X\\|U|=j\\|V|=|S|-j\end{array}}\!\!f(U)g(V)\,.
\end{split}
\end{equation}
Because $X$ ranges over all subsets of $S$, it follows that for any ordered 
pair $(U,V)$ of subsets of $S$ satisfying $|U|+|V|=|S|$, 
the term $f(U)g(V)$ occurs in the sum with sign $(-1)^{|S\setminus X|}$
exactly once for every $X$ satisfying $U\cup V\subseteq X\subseteq S$.
No other terms occur in the sum.
Thus, collecting the terms associated with each pair $(U,V)$ together,
the coefficient of $f(U)g(V)$ is $1$ if $U\cup V=S$ and $0$ otherwise.
Because $|U|+|V|=|S|$ and $U\cup V=S$ together imply $U\cap V=\emptyset$,
it follows that (\ref{eq:rank-convolution-inv}) and 
(\ref{eq:convolution2}) agree. In other words,
\begin{equation}
\label{eq:fast-convolution}
	(f*g)(S)
	=\sum_{X\subseteq S}(-1)^{|S\setminus X|}(\hat f\ostar\hat g)(|S|,X)\,.
\end{equation}

Given $f$ and $g$, we can now evaluate $f*g$ in $O(n^22^n)$ ring operations 
by first computing the fast ranked M\"obius transform 
of $f$ and $g$, then taking the convolution (\ref{eq:rank-convolution}) 
of the transforms $\hat f$ and $\hat g$, and inverting the result using 
fast ranked M\"obius inversion. This establishes Theorem~\ref{thm:main}.

\subsection{Variants and Extensions}
\label{sect:variants}

There are two immediate ways to relax the subset convolution
(\ref{eq:convolution2}).
First, the \emph{covering} product is defined for all $S\subseteq N$ by
\begin{equation}
\label{eq:covering-prod}
(f\covprod g)(S)=\sum_{\tiny\begin{array}{c}U,V\subseteq S\\U\cup V=S\end{array}} f(U)g(V)\,.
\end{equation}
Second, the \emph{packing} product is defined for all $S\subseteq N$ by
\begin{equation}
\label{eq:packing-prod}
(f\packprod g)(S)=\sum_{\tiny\begin{array}{c}U,V\subseteq S\\U\cap V=\emptyset\end{array}} f(U)g(V)\,.
\end{equation}

Given $f$ and $g$, the covering product (\ref{eq:covering-prod}) 
can be evaluated in $O(n2^n)$ ring operations by computing 
the M\"obius transforms $\hat f$ and $\hat g$, taking the 
elementwise (Hadamard) product 
$(\hat f\hat g)(X)=\hat f(X)\hat g(X)$ of the transforms, and 
inverting the result using fast M\"obius inversion.
Indeed, observe first that
\[
\sum_{X\subseteq S}(-1)^{|S\setminus X|}(\hat f\hat g)(X)=
\sum_{X\subseteq S}(-1)^{|S\setminus X|}\sum_{U,V\subseteq X}f(U)g(V)\,.
\]
Now, for each ordered pair $(U,V)$ of subsets of $S$, the coefficient of 
the term $f(U)g(V)$ is $1$ if $U\cup V=S$ and $0$ otherwise.
Thus,
\begin{equation}
\label{eq:covering-prod-fast}
(f\covprod g)(S)=\sum_{X\subseteq S}(-1)^{|S\setminus X|}(\hat f\hat g)(X)\,.
\end{equation}

Given $f$ and $g$, the packing product $f\packprod g$ can be evaluated 
in $O(n^22^n)$ ring operations by first computing the subset
convolution $f*g$ and then convolving the result with the vector
$\vec 1$ with all entries equal to $1$. 
Indeed, based on (\ref{eq:convolution2}) it is not difficult
to check that
\[
f\packprod g=f*g*\vec 1=f*\vec 1*g=\vec 1*f*g.
\]

Besides the immediate extensions (\ref{eq:covering-prod}) and 
(\ref{eq:packing-prod}), also somewhat more subtle variants are possible.
For example, using (\ref{eq:convolution2}) and (\ref{eq:covering-prod}),
define the \emph{intersecting covering product} for all 
$S\subseteq N$ by
\begin{equation}
\label{eq:covprodi}
(f\covprodi g)(S)=
\sum_{\tiny\begin{array}{c}U,V\subseteq S\\U\cup V=S\\U\cap V\neq\emptyset\end{array}} f(U)g(V)\,.
\end{equation}
A fast evaluation algorithm is now immediate from the
observation $f\covprodi g=f\covprod g-f*g$.
Also more precise control over the allowed intersection
cardinalities $|U\cap V|=\ell$ besides 
the $\ell=0$ ($f*g$) and $\ell>0$ ($f\covprodi g$) cases can 
be obtained by modifying (\ref{eq:rank-convolution}); however, 
we will not enter into detailed discussion.
Some further variations are possible by restricting the domain, e.g., to
any hereditary family of subsets of $N$; we omit the details.

\section{Model of Computation and The Choice of Ring}

Up to this point we have worked with an abstract ring $R$, 
and have considered only the number of ring operations 
(addition, subtraction, multiplication) required to carry out the
computations. 
To arrive at a more accurate analysis of the required computational
effort, we must choose a concrete ring $R$, fix a representation for 
its elements, and evaluate the required effort in a model that 
parallels the operation of an actual physical computer.
In what follows, the model of computation is the random
access machine with the restriction that arithmetic
operations (including comparison) are considered unit-time
only for constant-size integers. In this model, two
$b$-bit integers can be added, subtracted, and compared in 
$O(b)$ time, and multiplied in $O(b \log b \log \log b) = O^*(b)$
time \cite{Schonhage_and_Strassen_1971}.  
 
\subsection{Integer Sum--Product Ring.}

We prove Theorem~\ref{thm:intsumprod}.
We consider the subset convolution; similar argumentation
applies to the other variants in \ref{sect:variants}.  
By Theorem~\ref{thm:main}, we know that the subset convolution can be
computed in $O(n^2 2^n)$ ring operations. It is thus
sufficient to notice that any intermediate results, for which ring
operations are performed, are $O(n \log M)$-bit integers. To
see this, note first that the ranked M\"{o}bius
transform of an input function can be computed with integers
between $-M 2^n$ and $M 2^n$. Given this we note that the 
convolution of ranked transforms can be computed with $O(n \log M)$-bit 
integers. 
Finally, the ranked M\"{o}bius inversion is computed by
adding (and subtracting) $O(n \log M)$-bit integers $O(2^n)$ times.  

\subsection{Integer Max--Sum and Min--Sum Semirings.}

We prove Theorem~\ref{thm:intminsum}.
We consider the case of max--sum semiring; similar
argumentation applies to the min--sum semiring. 
Without loss of generality we assume that the range of the input
functions is $\{0, 1, \ldots, M\}$; otherwise, we may first add
$M$ to each value of both input functions, compute the
convolution, and finally subtract $2M$ to get the correct
output. 

Let $f$ and $g$ be the two input functions. Let 
$\beta = 2^n + 1$ and $M' = \beta^M$. 
Define  new mappings $f'$ and $g'$
from the subsets of $N$  to $\{0, 1, \ldots, M'\}$   by $f'
= \beta^f$ and $g' = \beta^g$. By Theorem~\ref{thm:intsumprod} 
we can compute the subset convolution $f' * g'$ over the integer
sum--product ring in $O^*(2^n \log M') = O^*(2^n M)$ time.
It remains to show that we can, for all $S \subseteq N$,
efficiently deduce the value of  $\max_{T \subseteq
S}\{f(T)+ g(S \setminus T)\}$ given the value of $\sum_{T
\subseteq S} f'(T) g'(S \setminus T)$.

We observe that, for all $S \subseteq N$, we have a 
polynomial representation 
\bes
	(f' * g')(S) = \sum_{T \subseteq S} \beta^{f(T) + g(S
	\setminus T)} = \alpha_0(S) + \alpha_1(S) \beta + \cdots
	+ \alpha_{2M}(S) \beta^{2M}\,,
\ees
where, due to the
choice of $\beta$, each coefficient $\alpha_r(S)$ is
uniquely determined and equals the number of subsets $T$ of $S$
for which $f(T) + g(S \setminus T)= r$. Thus, for each $S
\subseteq N$, we can find the largest $r$ for which
$\alpha_r(S) > 0$ in $O^*(M)$ time. This completes the
proof.

\section{Applications}
\label{se:applications}

\subsection{The Minimum Steiner Tree Problem}

The Steiner tree problem is a classical NP-hard problem.
Given an undirected graph $G = (V, E)$,  a weight $w(e) >
0$  for each edge $e \in E$, and a set of  vertices
$K\subseteq V$, the task is to find a subgraph $H$ of $G$
that connects the vertices in $K$ and has the minimum  total
weight $\sum_{e \in E(H)} w(e)$ among all such
subgraphs  of $G$. Because the edge weights are positive, an
optimal subgraph $H$  is necessarily a tree (a Steiner tree)
with leaves in $K$.

In accordance with the supposed model of computation, we
require that the edge weights are integers from $\{1, 2,
\ldots, M\}$. To simplify some expressions, we assume that
$M$ is a constant.

\subsubsection{Dreyfus--Wagner Recursion}
\label{sect:d-w-recursion}

Dreyfus and Wagner \cite{Dreyfus_and_Wagner_1972} 
discovered a beautiful dynamic programming algorithm for
finding a Steiner tree in $O^*(3^k n + 2^k n^2 + n
m)$ time,  where $n=|V|$, $m=|E|$, and $k=|K|$.

The key idea in the Dreyfus--Wagner algorithm is that 
a Steiner tree $H$ connecting a given subset of vertices $Y\subseteq V$ in $G$
has the following optimal decomposition property,
assuming $|Y|\geq 3$. 
For every $q\in Y$, there exists a vertex $p\in V$, 
a nonempty proper subset $D\subset Y\setminus\{q\}$, and 
a decomposition $E(H)=E(H_1)\cup E(H_2)\cup E(H_3)$ such that 
(a) $H_1$ is a Steiner tree connecting $\{p,q\}$ in $G$,
(b) $H_2$ is a Steiner tree connecting $\{p\}\cup D$ in $G$, and 
(c) $H_3$ is a Steiner tree connecting 
$\{p\}\cup(Y\setminus(D\cup\{q\}))$ in $G$.
(See \cite{Dreyfus_and_Wagner_1972} for a proof.)
Note that the decomposition may be degenerate, e.g., 
we can have $p=q$, implying that $H_1$ is empty.

The optimal decomposition property enables the following 
\emph{Dreyfus--Wagner recursion}. 
For a vertex subset $Y\subseteq V$, denote by $W(Y)$ the
total weight of a Steiner tree connecting $Y$ in $G$. 
To set up the base case, observe that
for $|Y|\leq 1$ the weight $W(Y)=0$ and 
for $|Y|=2$ the weight $W(Y)$ can be determined by a 
shortest-path computation based on the edge weights $w(e)$.
For $|Y|\geq 3$ the optimal decomposition property
implies that we have for all $q\in Y$ and $X=Y\setminus\{q\}$
the recursion
\be \label{eq:dw1}
  W(\{q\} \cup X) = \min\big\{W\bigl(\{p,q\}\bigr) + g_p\bigl(X\bigr)\,:\,
                              p\in V\big\}\,,
\ee
\be \label{eq:dw}
  g_p(X) =  \min
             \big\{ W\bigl(\{p\}\cup D\bigr) + 
                    W\bigl(\{p\}\cup (X\setminus D)\bigr)\,:\,
                    \emptyset\subset D\subset X\}\,.
\ee

The original problem can be solved by computing the weight
$W(K)$ via this recursion. A bottom-up
evaluation of $W(K)$ relying on dynamic programming takes the 
claimed $O^*(3^k n + 2^k n^2 + n m)$ time; first all-pairs shortest
paths are computed in $O^*(n^2 + nm)$ time (in $O(n^2 \log n
+ n m)$ basic operations) using, e.g., Johnson's algorithm
\cite{Johnson_1977}.  Once the values $W(\{p\}\cup Y)$ and
$g_p(Y)$  for all $Y\subset K$ and $p\in V$ have been
computed and stored, an actual Steiner tree that achieves
the optimal weight $W(K)$ is easy to construct by tracing
backwards a path of optimal choices in (\ref{eq:dw1}) and (\ref{eq:dw})
\cite{Dreyfus_and_Wagner_1972}; this costs only 
$O(2^k + kn)$ simple operations, that is, 
$O^*(2^k \log n + kn)$ time. 

\subsubsection{Expediting the Dreyfus--Wagner Recursion}
\label{sect:e-d-w}

We apply the fast subset convolution over the min--sum
semiring to expedite the evaluation of the Dreyfus--Wagner recursion
in (\ref{eq:dw}). However, we cannot simply replace (\ref{eq:dw}) by 
fast subset convolution as each $g_p(X)$ is defined in terms of other
values $g_r(Z)$, for $Z \subset X$ and $r \in V$, which need to be 
precomputed. To this end, we carry out the computations in a 
level-wise manner.

For each level $\ell=2,3,\ldots,k-1$ in turn, 
assume the value $W(\{q\}\cup X)$ has been computed and stored
for all $X \subset K$ with $|X|\leq\ell-1$ and $q \in V\setminus X$.  
To compute $g_p(X)$ for each $p \in V$ and $X \subset K$ with 
$|X| = \ell\geq 2$, define the function $f_p$ for all $X\subseteq K$ by
\begin{equation}
\label{eq:d-w-levelfunc}
f_p(X)=
\begin{cases}
W(\{p\}\cup X)  & \text{if $1\leq|X|\leq\ell-1$}\,,\\
\infty          & \text{otherwise}\,.
\end{cases}
\end{equation}
Here we let $\infty$ in (\ref{eq:d-w-levelfunc}) denote
an integer that is sufficiently large to exceed the weight 
of any tree in $G$; for example, $(n-1)M+1$ suffices.
Applying the subset convolution over the min--sum semiring,
it is now immediate from (\ref{eq:dw}) and (\ref{eq:d-w-levelfunc}) 
that $g_p(X)=(f_p*f_p)(X)$ holds for all $X\subseteq K$ with $|X|\leq\ell$. 
Thus, by Theorem~\ref{thm:intminsum}, we can compute $g_p(X)$ 
for all $p \in V$ and $X\subset K$ with $|X| = \ell$ 
using $n$ evaluations of the subset convolution with integers
bounded by $nM$, which leads to $O^*(2^k n^2)$ total time;
note that the $O^*$ notation hides a factor of $k^3$.
In fact, we can do even better and save a factor of $k$ 
by replacing the subset convolution with the covering product over 
the min--sum semiring.
To see this, observe that because $W(Z)\leq W(Y)$ holds
whenever $Z\subseteq Y\subseteq V$, we have that 
(\ref{eq:dw}) can also be computed as 
$g_p(X) = (f_p \covprod f_p)(X)$, that is, 
\[
g_p(X) =  \min
           \big\{ W\bigl(\{p\}\cup T\bigr) + 
                  W\bigl(\{p\}\cup U\bigr)\,:\,
                  \emptyset\subset T,\!U\subset X,\,\ T\cup
				  U=X\}\,.
\]
Once the values $g_p(X)$ have been computed, it is easy to
compute $W(\{q\}\cup X)$ for all  
$X\subset K$ and $q\in V\setminus X$ with $|X| = \ell$
in $O^*\big({k \choose \ell}n^2\big)$ time using (\ref{eq:dw1}).
Computing the above steps for all levels $\ell = 2, 3,
\ldots, k-1$ takes $O^*\bigl(2^k n^2 + nm\bigr)$ total
time, including the time needed for computing all-pairs
shortest paths. Finally, 
a Steiner tree can be constructed within the same time bound
(see \S\ref{sect:d-w-recursion}).
We have thus established the following theorem, which we 
state in a form without the assumption that $M$ is constant.

\begin{Thm}
\label{thm:steiner}
The minimum Steiner tree problem with edge weights in\/
$\{1,2,\ldots,M\}$ can be solved in 
$O^*(2^k n^2 M + n m \log M)$ time.
\end{Thm}

\subsection{A Rooted Tree Model for Signaling Pathways}

Scott, Ideker, Karp, and Sharan \cite{Scott_etal_2005} 
consider various models for signaling pathways in protein 
interaction networks.
One of the two more general models they introduce is based
on rooted trees, and leads to the following network problem.
Given an undirected graph $G = (V, E)$, a weight
$w(e)$ for each edge $e \in E$, 
a vertex subset $I\subseteq V$, and a positive integer $k$, 
the task is to find for each vertex $v \in V$ a tree of 
the minimum total weight among all $k$-vertex subtrees 
in $G$ that are rooted at $v$ and in which every leaf belongs to $I$.

Scott et al.\ \cite{Scott_etal_2005} apply the 
color coding method of Alon, Yuster, and Zwick \cite{Alon_etal_1995},
which proceeds by carrying out a sequence of randomized trials.
In each trial, every vertex $v$ is given independently 
and uniformly at random a color $c(v)\in\{1,2,\ldots, k\}$, 
and the following subtask is solved:
for each vertex $v \in V$ and subset 
$S\subseteq\{1,\ldots,k\}$ that contains $c(v)$, 
find a minimum-weight subtree with $|S|$ vertices that is 
(a) rooted at $v$, 
(b) contains a node of each color in $S$, and 
(c) in which every leaf belongs to $I$. 
Scott et al.\ give the following recurrence
for the associated minimum weight, denoted by $W(v, S)$:
\bes
  W(v, S) = \min\bigl\{A(v, S),\,B(v, S)\bigr\}\,,
\ees
where
\[
\begin{split}
  A(v, S) &= \min
     \bigl\{W(u, S\setminus\{c(v)\}) + w(u, v)\,:\,
            c(u) \in S \setminus \{c(v)\}\bigr\}\,,\\
  B(v, S) &= \min
      \bigl\{W(v, T) + W(v, U)\,:\,
             T \cap U  = \{c(v)\},\, T \cup U = S\bigr\}\,,
\end{split}
\]
with 
$W(v, \{c(v)\}) = 0$ if $v \in I$ and 
$W(v, \{c(v)\}) = \infty$ otherwise. 
A direct evaluation of this recurrence can be carried out
in $O^*(3^k m)$ time \cite{Scott_etal_2005}, where $m=|E|$.

Armed with fast subset convolution, we can
speed up the evaluation of the recurrence to $O^*(2^k m)$ time, 
assuming that the edge weights are small integers. Namely, 
proceeding simultaneously for all sets $S$ of a given cardinality,
the computation of $B(v, S)$ can be reduced to subset convolution over 
the integer min--sum semiring; the transformation is analogous to the 
one used in \S\ref{sect:e-d-w}, so we omit details.

Scott et al.\ \cite{Scott_etal_2005} also consider 
a different model based on two-terminal series-parallel graphs. 
In this case, too, the original $O^*(3^k n^2)$ algorithm can be 
accelerated to an $O^*(2^k n^2)$ algorithm by using fast subset convolution.

\subsection{Partitioning Problems and Extensions}

Consider the generic problem of partitioning an $n$-element
set $N$ into $k$ disjoint subsets that each satisfy some
desired property specified by an indicator function $f$ on
the subsets of $N$. Given, $N$, $k$, and $f$ as input, the
task is to decide whether there exists a partition $\{S_1,S_2,
\ldots, S_k\}$ of $N$ such that $f(S_c) = 1$ for each $c =
1, 2, \ldots, k$. Many classical graph partitioning problems
are of this form. For example, in graph coloring $f(S) = 1$
if and only if $S$ is an independent set in the input graph
with the vertices $N$. Likewise, in domatic partitioning
$f$ is the indicator of dominating sets. 

Recently we \cite{Bjorklund_and_Husfeldt_2006,
Koivisto_2006} discovered two different algorithms that
solve the generic partitioning problem using the principle
of inclusion and exclusion in $O^*(2^n)$ time, provided that
$f(S)$ can be evaluated for all $S \subseteq N$ in
$O^*(2^n)$ total time. Using fast subset convolution we
obtain yet another $O^*(2^n)$ algorithm. Indeed, we observe
that the number of valid partitions of $N$ is given by
$f^{*k}(N)$, where
\bes
	f^{*k} = \underbrace{f * f * \cdots *
	f}_{k \textrm{ times}}\,.
\ees 
Thus, we can count the valid partitions by $k-1$ 
subset convolutions, or even better, in $O(\log k)$
convolutions by using the doubling trick.

What is more, we can solve considerable extensions of
partitioning problems within the same runtime bound. For
example, we can find a maximal $k$-colorable induced
subgraph (in fact, all such subgraphs)  in $O^*(2^n)$ time
by computing $f^{*k}(S)$ for all vertex subsets $S \subseteq
N$. In a similar fashion, but using the packing product, we
can decide whether the input graph $G$ contains $k$ disjoint
cliques each of size at least $\ell$ in $O^*(2^n)$ time: we
check if $f^{\packprod k}(N) > 0$, where $f(S) = 1$ if 
$S$ is a clique in $G$ with $|S| \geq \ell$, and $f(S)=0$ otherwise.

Fast subset convolution allows us to solve not only flat
partitioning problems but also {\em hierarchical}
partitioning problems in $O^*(2^n)$ time. Consider, for
example, a branching process that partitions the ground set
$N$ in a tree-structured manner, as follows. With
probability $\alpha$, a node $S \subseteq N$ is split
uniformly at random into
two proper subsets $T \subset S$ and $S \setminus T \subset
S$, which are then further
partitioned recursively; with the remaining probability
$1-\alpha$, the branching terminates at $S$, and $S$ becomes a
leaf of the tree. With each possible leaf $L \subseteq N$ we
associate a number $f(L)$, and by $g(N)$ we denote the
expected value of the product of $f(L)$ over all leafs of
the (random) tree. Then $g(N)$ can be solved through 
a recursion for $S \subseteq N$:
\bes
	g(S) = (1-\alpha) f(S) + \alpha \cdot \frac{1}{2^{|S|}
	- 2}
	\sum_{\emptyset \subset T \subset S} g(T) g(S \setminus
	T)\,.
\ees
Using fast subset convolution we can compute
$g(S)$ for all $S \subseteq N$ in a total of $O^*(2^n)$
arithmetic operations.

\subsection{Spanning Problems in Hypergraphs}

We conclude this section by illustrating more subtle
applications to two NP-hard hypergraph problems
(see \cite{Polzin_and_Daneshmand_2003,Warme_1998}).
We begin by recalling the appropriate hypergraph terminology.
A \emph{hypergraph} is a pair $\mathcal{H}=(V,\mathcal{E})$,
where $V$ is a finite set and $\mathcal{E}$ is a set consisting
of subsets of $V$. A hypergraph $\mathcal{J}=(W,\mathcal{F})$
is a \emph{subhypergraph} of $\mathcal{H}$ if $W\subseteq V$ and
$\mathcal{F}\subseteq\mathcal{E}$.
A subhypergraph is \emph{spanning} if $V=W$.
A \emph{path} in a hypergraph $\mathcal{H}$ is 
a sequence $(x_1,E_1,x_2,E_2,\ldots,E_{\ell},x_{\ell+1})$
such that 
(a) $x_1,x_2,\ldots,x_{\ell+1}\in V$ are all distinct,
(b) $E_1,E_2,\ldots,E_\ell\in\mathcal{E}$ are all distinct,
and 
(c) $x_i,x_{i+1}\in E_i$ for all $i=1,2,\ldots,\ell$. 
A path \emph{joins} $x_1$ to $x_{\ell+1}$. 
A hypergraph is \emph{connected} if for all distinct 
$x,y\in V$ there exists a path joining $x$ to $y$. 
A connected hypergraph is a \emph{tree} if 
for all distinct $x,y\in V$ the path joining $x$ to $y$ is unique.

The \emph{minimum connected spanning subhypergraph} (MCSH) problem 
asks, given a hypergraph $\mathcal{H}=(V,\mathcal{E})$ 
and a weight $w(E)>0$ for each hyperedge $E\in\mathcal{E}$,
to produce a connected spanning subhypergraph of $\mathcal{H}$
that has the minimum total weight, or to assert that none exists.
The \emph{minimum spanning tree} (MSTH) problem is otherwise
similar to the MCSH problem, but in addition it is required 
that the subhypergraph must be a tree.

Assuming that $w(E)\in\{1,2,\ldots,M\}$ for all $E\in\mathcal{E}$,
both the MCSH problem and the MSTH problem can be solved in
time $O^*(2^nM)$ using variants of the fast subset convolution
over the min--sum semiring, where $n=|V|$.
Indeed, define the function $f$ for all $E\subseteq V$ by
\[
f(E)=\begin{cases}
w(E)   & \text{if $E\in\mathcal{E}$}\,,\\
\infty & \text{otherwise}\,.
\end{cases}
\]
To solve the MCSH problem, we employ the 
intersecting covering product (\ref{eq:covprodi}).
Define the $k$th power of the intersecting covering product
for all $k=2,3,\ldots$ by
\[
f^{\covprodi k}=f\covprodi \bigl(f^{\covprodi(k-1)}\bigr),\quad
f^{\covprodi 1}=f.
\]
Here the order in which the products are evaluated is relevant
because the intersecting covering product is \emph{not} associative.
Now observe that 
(a) a MCSH can be constructed by augmenting a connected 
subhypergraph of $\mathcal{H}$ one hyperedge at a time, and
(b) at most $n-1$ hyperedges occur in a MCSH of $\mathcal{H}$.
Thus, $f^{\covprodi k}(V)<\infty$ is the minimum weight of a connected 
spanning subhypergraph of $\mathcal{H}$ consisting of $k$ hyperedges.
By storing the functions $f^{\covprodi k}$ for each $k=1,2,\ldots,n-1$, 
the actual MCSH can be determined by tracing back the computation 
one edge at a time.
To solve the MSTH problem, replace the intersecting covering 
product (\ref{eq:covprodi}) with an intersecting covering 
product that in addition requires the cardinality of the 
intersection to be exactly $1$; such a product can be obtained 
by a minor modification of (\ref{eq:rank-convolution}).

\section*{Acknowledgments}
\noindent
This research was supported in part by the Academy of
Finland, Grants 117499 (P.K.) and 109101 (M.K.).

%%%%%%%%%%%%%%%%%%%%%%%%%%%%%%%%%%%%%%%%%%%%%%%%%%%%%%%%%%%%%%%% References %%%

\end{document}